\PassOptionsToPackage{svgnames}{xcolor}

\documentclass[]{aastex701}
\usepackage{graphicx}%
\usepackage{multirow}%
\usepackage{amsmath,amssymb,amsfonts}%
\usepackage{amsthm}%
\usepackage{mathrsfs}%
\usepackage[title]{appendix}%
\usepackage{xcolor}%
\usepackage{textcomp}%
\usepackage{booktabs}%
\usepackage{algorithm}%
\usepackage{algorithmicx}%
\usepackage{algpseudocode}%
\usepackage{listings}%

\newcommand{\HeI}{\ion{He}{i} 10830 \AA}

\DeclareRobustCommand{\ion}[2]{%
\relax\ifmmode
\ifx\testbx\f@series
{\mathbf{#1\,\mathsc{#2}}}\else
{\mathrm{#1\,\mathsc{#2}}}\fi
\else\textup{#1\,{\mdseries\textsc{#2}}}%
\fi}

\begin{document}

\title{Reconstructing Synthetic SDO/AIA 193~\AA{} EUV Images from \HeI{} Observations with Diffusion Model Translator}

\correspondingauthor{Bo Shen}
\email{bo.shen@njit.edu}

\author{Marco Marena}
\affiliation{Department of Mechanical and Industrial Engineering, New Jersey Institute of Technology}
\email{mm243@njit.edu}

\author{Qin Li}
\affiliation{Department of Physics, New Jersey Institute of Technology}
\email{ql47@njit.edu}

\author{Haimin Wang}
\affiliation{Department of Physics, New Jersey Institute of Technology}
\email{haimin.wang@njit.edu}

\author{Haodi Jiang}
\affiliation{Department of Computer Science, Sam Houston State University }
\email{hxj024@shsu.edu}

\author{Prajwal Shah}
\affiliation{Department of Computer Science, New Jersey Institute of Technology}
\email{ps2249@njit.edu}

\author{Bo Shen}
\affiliation{Department of Mechanical and Industrial Engineering, New Jersey Institute of Technology}
\affiliation{Department of Data Science, New Jersey Institute of Technology}
\email{bo.shen@njit.edu}

\begin{abstract}
Routine full-disk EUV imaging has been available only since the modern era, such as SOHO and SDO. To extend EUV coronal context into earlier periods, we leverage the multi-decade availability of full-disk \HeI{} observations, whose absorption is modulated by coronal irradiance and magnetic topology and is widely used as a proxy for open-field regions. We present a diffusion-based conditional image translation framework, Coronal Hole-aware Diffusion Model Translator (CH-aware DMT), to reconstruct synthetic SDO/AIA~193~\AA{} EUV images from \HeI{} inputs. The model is trained on temporally co-aligned SOLIS \HeI{} and AIA~193~\AA{} pairs spanning 2011--2015 using a month-based split, where January--October are used for training, November is used for validation, and December for testing. On the held-out test set, the reconstructions preserve dominant full-disk EUV morphology (CC=0.92) and recover CH-related low-intensity structure (CC=0.84). We further assess historical applicability by (1) comparing reconstructed AIA~193~\AA{} morphology with SOHO/EIT~195~\AA{} over 2005--2015; (2) comparing reconstructed AIA~193~\AA{} images generated from KPVT \HeI{} inputs against Yohkoh/SXT soft X-ray observations; and (3) evaluating long-term reconstructed disk-integrated emission statistics against observational EUV series and independent solar activity proxies (sunspot number and F10.7 radio flux over 1974--2015). These results indicate that CH-aware DMT conditioned on \HeI{} can provide a physically plausible synthetic AIA~193~\AA{} coronal proxy for historical studies, supporting multi-decade analyses of large-scale coronal evolution before the direct EUV imaging was available.
\end{abstract}

\keywords{Solar corona --- EUV --- SDO/AIA 193~\AA{} --- \HeI{} --- Coronal holes --- Diffusion models --- Image-to-image translation}

\section{Introduction}\label{sec:intro}

The Extreme Ultraviolet (EUV) 193~\AA{} channel of the Solar Dynamics Observatory’s Atmospheric Imaging Assembly (SDO/AIA) provides a cornerstone diagnostic for the hot solar corona due to its high cadence, full disk coverage, and fine spatial sampling \citep{Pesnell2012,Lemen2012}. The 193~\AA{} passband is dominated by coronal iron emission, primarily Fe \textsc{xii} formed near $\sim$1.3~MK, with additional high temperature contributions (e.g., Fe \textsc{xxiv}) becoming relevant during flares \citep{ODwyer2010}. In typical conditions, active regions (ARs) appear bright in AIA 193~\AA{} due to enhanced emission from dense, heated coronal loops, whereas coronal holes (CHs) manifest as dark, low-intensity patches \citep{Cranmer2009,ODwyer2010}. These dark regions correspond to volumes of reduced temperature and density associated with open magnetic flux that channels the fast solar wind \citep{Krieger1973}. Because the emissivity of highly ionized coronal iron is strongly diminished in CH environments \citep{Nolte1976}, AIA 193~\AA{} imagery is particularly effective for delineating CH boundaries and tracking their evolution through the solar cycle. Beyond CH mapping, the 193~\AA{} channel captures a wide range of coronal phenomena—from quiescent loop arcades to eruptions supporting investigations of the corona--heliosphere connection and space-weather drivers.

While the EUV imager, SDO/AIA, has only been available since 2010, the solar community has long sought reliable historical proxies for coronal structure. One of the most important of these is the neutral-helium infrared triplet at 10830 \AA{}. Although formed in the upper chromosphere, the \HeI{} line is unusually sensitive to the state of the overlying corona because its population mechanisms depend on coronal irradiance and the thermodynamic coupling of the atmosphere \citep{AndrettaJones1997}. In CH regions, \HeI{} absorption is typically reduced, making CHs appear relatively bright in spectroheliograms \citep{Nolte1976}. This behavior is commonly interpreted in terms of reduced coronal EUV irradiance above CHs and the resulting shift in the balance of processes that populate the helium triplet, leading to weaker absorption (higher intensity) over open-flux regions. Since early recognition of the diagnostic value of \HeI{} for CH identification, synoptic observations have provided an essential long-term perspective on CH morphology across multiple solar cycles, including eras that predate routine EUV imaging \citep{HarveyRecely2002}. This capability is especially valuable for reconstructing coronal conditions and solar-wind source regions in historical intervals, well before modern EUV telescopes \citep{Krieger1973}. While \HeI{} is a valuable CH proxy, it is formed in the upper chromosphere. It primarily provides an absorption-based view that is modulated by coronal irradiance and thermodynamic coupling, rather than a direct coronal-emissivity diagnostic. Please note that \HeI{} does not directly measure coronal emission. It provides a chromospheric absorption response that is influenced by coronal conditions and therefore does not contain the same information as an EUV coronal image.
In contrast, AIA~193~\AA{} directly samples coronal Fe~\textsc{xii} emission near $\sim$1--1.5~MK under non-flaring conditions, providing a physically interpretable coronal context for both low-emission domains and bright coronal loop systems \citep{ODwyer2010,Lemen2012}. Extending an AIA~193~\AA{} product, therefore, enables historical analyses and cross-era comparisons that are naturally formulated in the EUV coronal height (e.g., with EIT~195~\AA{} and soft X-ray references)

Full-disk \HeI{} observations provide a historically longer record than space-based EUV imagery. The NSO/KPVT \HeI{} record extends back to the 1970s, while NSO/SOLIS provides continued full-disk \HeI{} observations beginning in the 2000s. In contrast, SOHO/EIT 195~\AA{} observations begin in the mid-1990s, and SDO/AIA 193~\AA{} observations begin in 2010. This longer temporal baseline makes \HeI{} a valuable historical proxy for extending coronal-hole diagnostics into periods before the SDO era. Our goal is therefore to use full-disk \HeI{} observations as a physically motivated proxy to generate synthetic coronal imagery in the AIA 193~\AA{} band, extending EUV CH diagnostics into the pre-SDO era. Reconstructing historical coronal EUV emission from chromospheric proxies, however, is a non-trivial spectral translation problem. The mapping from \HeI{} absorption to EUV coronal emission is nonlinear and mediated by the 3D thermodynamic structure and radiative coupling of the solar atmosphere \citep{AndrettaJones1997,Cranmer2009}. Traditional approaches have therefore relied on correlations, heuristic thresholds, or manual co-registration to identify common CH regions in different diagnostics \citep{Nolte1976,HarveyRecely2002}. While effective for generating CH masks and synoptic boundary maps, these strategies are not designed to synthesize realistic, high-fidelity EUV intensity images at AIA resolution and radiometric complexity \citep{ODwyer2010}.

Recent advances in data-driven modeling have opened a new pathway: learning the translation between solar observables directly from paired observations. In the broader machine-learning (ML) literature, conditional generative models such as conditional Generative Adversarial Networks (cGANs) and image-to-image (I2I) translation frameworks have become standard tools for learning mappings between different domains \citep{Goodfellow2014,Isola2017, yuan2020attribute,yuan2023dde,ghasemi2024dcg}. Within solar physics, a growing body of work has adopted artificial intelligence (AI) to infer otherwise unobserved solar quantities and to generate proxy observables when direct measurements are sparse or unavailable. A major motivation is that many solar data products are not co-observed continuously across missions (or across historical eras), so learning a mapping between modalities can provide physically meaningful ``synthetic'' views that extend the effective observational record.

Early and influential examples include the use of cGANs and related encoder--decoder models to translate between magnetic and radiative diagnostics \citep{Dash2021,Jarolim_2025}. In a prominent series of studies, it has been demonstrated that deep generative models can recover magnetic-structure information from EUV context and, conversely, can generate EUV/UV emission proxies from magnetic maps. For example, farside magnetograms can be generated from STEREO/EUVI imagery using cGAN-based approaches, enabling continuous tracking of AR magnetic evolution when direct farside magnetograms are unavailable \citep{Kim2019FarsideNatAstron,Jeong2022AISFM,Sun2022DynamicMagSeq,Dannehl2024EUVtoMagnetogram}. Furthermore, neural translation models that map photospheric magnetograms to multi-channel UV/EUV emission have been shown to recover large-scale coronal morphology and AR structure, supporting the idea that learned cross-modal mappings can preserve diagnostically relevant patterns across disparate solar observables \citep{Park2019MagToEUV,Salvatelli2022EUVLimits,doSantos2021AIAAutocal}. Recent work has also highlighted the importance of controlling hallucinations and physically inconsistent features in cross-modal translation. It has shown benefits from non-adversarial and modulation-based conditioning strategies \citep{Sayez2025Hallucination}.

Related efforts have also targeted chromospheric--coronal proxies. In particular, deep learning has been used to generate \HeI{} imagery from combinations of AIA EUV channels, effectively synthesizing chromospheric helium diagnostics with higher cadence and fewer gaps than the original ground-based \HeI{} record \citep{Son2021HeIfromAIA}. Building on this general paradigm, recent work has shown that AI-based image translation can reconstruct synthetic \HeI{} full-disk images directly from H$\alpha$ observations (e.g., using a cGAN/pix2pixHD-style framework), helping mitigate the limited historical availability of \HeI{} relative to the much longer H$\alpha$ archive \citep{Marena2025HeIfromHa}. AI-generated EUV products have also been used as inputs for downstream coronal diagnostics, such as differential emission measure (DEM) inference in settings where only a subset of EUV channels is observed \citep{Youn2025DEMAIGenerated}.

These results collectively motivate the approach taken here: treating long-running ground-based or synoptic diagnostics (e.g., \HeI{}) as historical proxies from which modern coronal observables can be synthesized. In the present work, this motivates learning a mapping from \HeI{} to EUV coronal structure (synthetic AIA~193~\AA{} output), and validating the realism of the synthesized products against the best available coronal references in each era. During overlap periods, these references include direct EUV observations such as SOHO/EIT~195~\AA{} and SDO/AIA~193~\AA{}; for earlier intervals, we include alternative coronal diagnostics such as soft X-ray observations and independently segmented CH products. Beyond image-level similarity alone, we are therefore interested in whether the reconstructed products preserve physically meaningful large-scale behavior, including long-term emission trends and Carrington-rotation synoptic structure. Nonetheless, common challenges remain: preserving fine structures, avoiding synthetic artifacts, and ensuring robust performance on rare or extreme events that are underrepresented in training distributions.

Motivated by these needs, we pursue diffusion-based image translation, building on Diffusion Model Translators (DMTs) \citep{Xia2025DMT} to generate synthetic AIA~193~\AA{} EUV images conditioned on \HeI{} observations. Diffusion models have become a preferred approach for image generation and I2I translation because they consistently achieve higher fidelity and better mode coverage than adversarial models in standard benchmarks \citep{DhariwalNichol2021DMBeatGANs}, and because diffusion-based I2I frameworks have been shown to outperform strong GAN and regression baselines across multiple translation tasks without requiring task-specific losses or architectural customization \citep{Saharia2022Palette}. In solar-physics imaging, diffusion models have also been shown to be effective generative priors for producing realistic synthetic SDO/AIA outputs, supporting their suitability for coronal proxy generation \citep{Ramunno2024DDPM}.

The overarching objective is to leverage the multi-decade availability of synoptic and full-disk \HeI{} imagery to reconstruct EUV coronal structure across earlier solar cycles, while using diffusion-based generation to improve image fidelity and reduce spurious artifacts. By conditioning a diffusion model on \HeI{} maps, we aim to learn the subtle, nonlinear mapping between chromospheric absorption signatures and coronal EUV emission, with particular emphasis on physically plausible large-scale morphology and temporal consistency. Because direct AIA~193~\AA{} ground truth (GT) does not exist before 2010, evaluation must rely on the most appropriate coronal reference available in each historical interval, emphasizing morphological consistency of low-emission domains and large-scale coronal structure rather than strict photometric agreement. For this purpose, SOHO/EIT~195~\AA{} provides the closest historical proxy to AIA~193~\AA{} for qualitative and trend-based comparisons of coronal morphology \citep{Delaboudiniere1995}. For earlier intervals without routine full-disk EUV observations, an independent coronal benchmark is required to assess whether the synthesized EUV structures remain physically plausible beyond the training distribution. We therefore leverage full-disk soft X-ray imaging from the Yohkoh/Soft X-ray Telescope (SXT; \citealt{Tsuneta1991}), which overlaps the KPVT \HeI{} synoptic record in the early 1990s. Although SXT has a different temperature response than the AIA~193~\AA{} passband, coronal holes and other low-emission structures manifest in both EUV and soft X-ray diagnostics, making SXT a suitable reference for morphology-based checks of large-scale coronal structure during 1991--1992 \citep{Kahler1983,Cranmer2009}. 


\section{Methods}\label{sec:methods}

\subsection{Data for Model Development}\label{sec:data}

We construct a paired dataset spanning November 2011 to December 2015 by combining full-disk \HeI{} observations from SOLIS with temporally co-aligned SDO/AIA 193~\AA{} images \citep{Lemen2012, NSOSOLISData, Henney2006}. To reduce temporal evolution effects while maximizing daily coverage, we select at most one paired observation per day and associate each \HeI{} image with the closest in time AIA 193~\AA{} acquisition within a tolerance of $\pm$\,1 minute.

In total, the dataset contains \textbf{829} paired full-disk samples, with \textbf{697} used for training, \textbf{75} for validation, and \textbf{57} for testing.  The data covered a variety of solar activity levels to make the training robust. Because neighboring daily solar images are highly correlated in morphology, a random day-level split would risk optimistic performance estimates by allowing the model to train on samples that are only a few days away from those used for evaluation. To reduce this short-timescale leakage, we reserve all images from \textbf{November} for validation and all images from \textbf{December} for testing, while using the remaining months from \textbf{January} to \textbf{October} for training, and repeat this rule consistently for each year in the 2011--2015 interval.

We also chose not to split the data by year. Over a relatively short five-year interval, a year-based split would be strongly affected by solar-cycle variability, leading to uneven distributions of key solar structures across the train, validation, and test sets. In particular, the frequency and complexity of active regions and coronal holes vary substantially from year to year, so holding out an entire year could produce evaluation sets with feature distributions that are not well represented during training. The month-based split provides a better compromise: it preserves temporal separation at the scale needed to avoid leakage from consecutive dates, while maintaining a more balanced distribution of coronal hole and activity-related features across all splits. To quantify the latter, we use the daily total sunspot number from the SILSO Version 2.0 record as a proxy for overall solar activity \citep{SILSO_Sunspot_Number}. As shown in Table~\ref{tab:dataset_ch_sunspot_stats}, the training, validation, and test sets contain comparable fractions of the total detected CH (\(84.95\%\), \(8.59\%\), and \(6.46\%\), respectively) and of the total sunspot-activity proxy (\(82.28\%\), \(10.58\%\), and \(7.14\%\), respectively), closely matching the corresponding image fractions of \(84\%\), \(9\%\), and \(7\%\). This supports the use of a month-based split as a practical compromise between temporal separation and a reasonably balanced distribution of morphology and activity level across the three subsets.

\begin{table}[ht]
\centering
\caption{Dataset statistics for the SDO/AIA 193~\AA{} splits. CH \% of total is computed from the total number of detected coronal holes across all splits. Sunspot \% of total is computed from the total daily sunspot number summed over all dates in each split, using the SILSO Version 2.0 daily total sunspot number record as a proxy for solar activity.}
\begin{tabular}{lrrrrrr}
\hline
Dataset & Images & \% of Total & CH Count & CH \% of Total & Sunspot Sum & Sunspot \% of Total \\
\hline
Training (January--October) & 697 & 84\% & 2353 & 84.95\% & 63538 & 82.28\% \\
Validation (November)   & 75  & 9\%  & 238  & 8.59\%  & 8172  & 10.58\% \\
Test (December)  & 57  & 7\%  & 179  & 6.46\%  & 5511  & 7.14\% \\
\hline
\end{tabular}
\label{tab:dataset_ch_sunspot_stats}
\end{table}

All data are preprocessed within a uniform pipeline designed to handle heterogeneous full-disk solar observations across instruments. FITS files are read together with their header metadata so that all images can be placed in a consistent solar coordinate framework. We standardize the orientation of each observation by rotating the images so that solar north is up, following standard solar image coordinate conventions \citep{Thompson2006}. We then isolate the solar disk using header-derived geometric information, such as the disk center \((x_0,y_0)\) and apparent solar radius \(R_\odot\) in image coordinates, so that off-disk background does not contribute to the learning signal. After disk isolation, each image is resized to \(256\times256\) pixels. This resolution is sufficient for the present study, whose main objective is to reconstruct large-scale coronal morphology and its temporal evolution, especially the structure and transitions of coronal holes, rather than to recover the finest small-scale coronal texture.

For intensity handling, AIA 193~\AA{} images are normalized consistently across the dataset using a fixed intensity scaling defined by prescribed \(v_{\min}\) and \(v_{\max}\). An analogous normalization is applied to the \HeI{} inputs so that the model sees a comparable dynamic range across varying observing conditions.

\subsection{Diffusion Model Translator}\label{sec:dmt}
\begin{figure*}[!htbp]
    \centering
    \includegraphics[width=\textwidth]{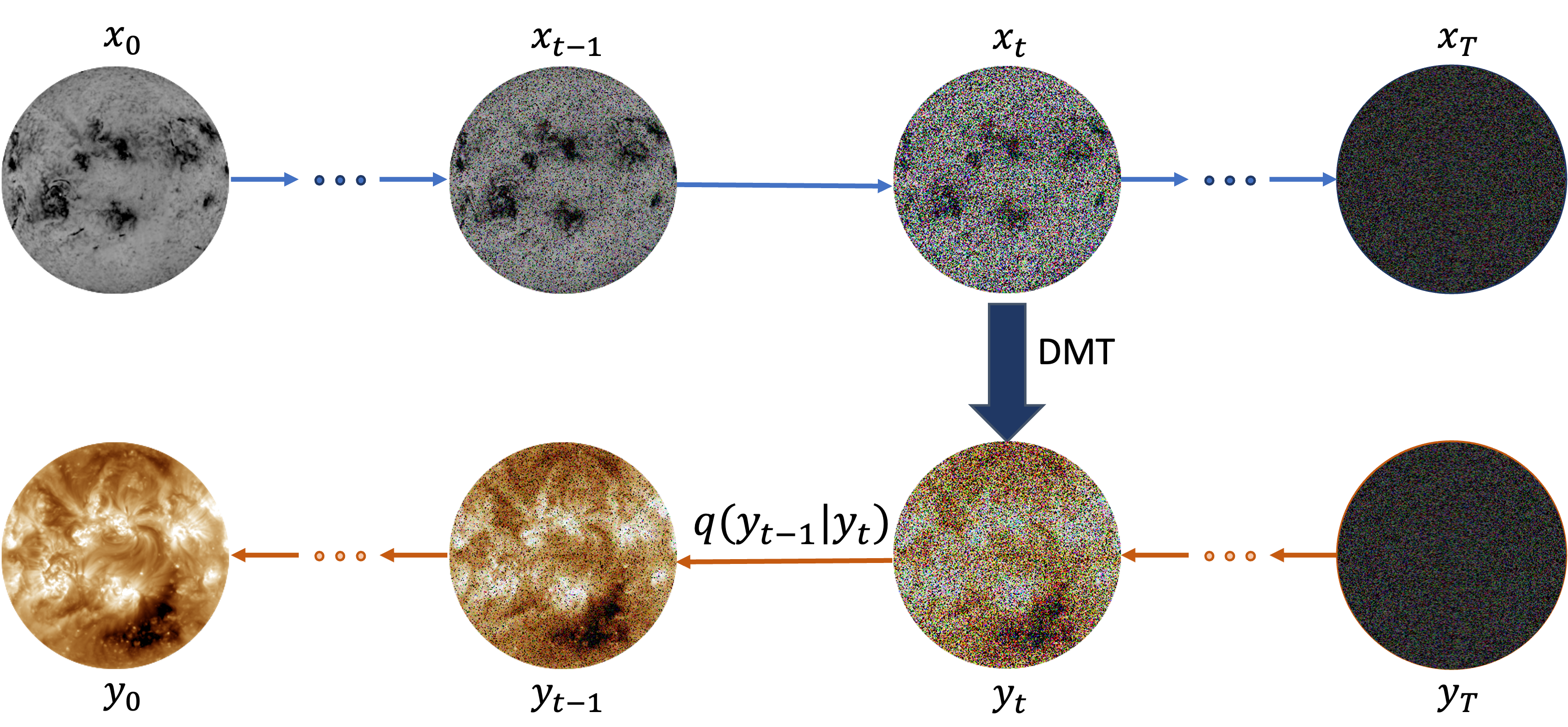}
    \caption{\textbf{Diffusion Model Translator (DMT) for conditional EUV reconstruction.}
    Top: The input $x_0$ (\HeI{}) is progressively noised by a forward diffusion process to obtain $x_t, t=1,\cdots, T$.
    Bottom: In parallel, the target domain $y$ (AIA~193~\AA{}) is generated by a learned reverse denoising chain from $y_T$ to $y_0$ via $q(y_{t-1}\!\mid y_t)$, conditioned at each timestep on the corresponding \HeI{} latent variable $x_t$ through the DMT module.}
    \label{fig:dmt_model}
\end{figure*}

We adopt the \emph{Diffusion Model Translator} (DMT) framework \citep{Xia2025DMT} to perform conditional image-to-image translation from a source observation ($x_0$, the full-disk \HeI{}) to a target-domain image ($y_0$, SDO/AIA~193~\AA). As summarized in Figure~\ref{fig:dmt_model}, DMT decomposes conditional generation into three stages: (i) a \emph{shared} forward diffusion that maps both domains to an intermediate noisy representation, (ii) a lightweight learned translator that performs the domain transfer \emph{at a single timestep}, and (iii) a reverse denoising process where a DDPM is trained on the target domain to convert the translated latent back to the clean target image.

\textbf{Forward diffusion.}
Let $x_0 \sim \mathcal{D}_x$ and $y_0 \sim \mathcal{D}_y$ be paired samples. Following denoising diffusion probabilistic models (DDPMs; \citealt{Ho2020DDPM}), we define a forward Markov diffusion process with variance schedule $\{\beta_t\}_{t=1}^T$ and $\alpha_t=1-\beta_t$, $\bar{\alpha}_t=\prod_{i=1}^t \alpha_i$. The $t$-step marginal has a closed form:
\begin{equation}
q(x_t\mid x_0)=\mathcal{N}\!\left(x_t;\sqrt{\bar{\alpha}_t}\,x_0,\,(1-\bar{\alpha}_t)\mathbf{I}\right),
\end{equation}
and analogously for $q(y_t\mid y_0)$. In practice, we encode the condition image by sampling $x_t$ at a pre-selected timestep $t^\star$:
\begin{equation}
x_{t^\star}=\sqrt{\bar{\alpha}_{t^\star}}\,x_0+\sqrt{1-\bar{\alpha}_{t^\star}}\,\epsilon,\quad \epsilon\sim\mathcal{N}(0,\mathbf{I}).
\end{equation}
This ``shared'' diffusion encoder is the mechanism by which DMT injects the conditioning information without requiring explicit conditioning at every reverse step.

\textbf{Single-step domain transfer with a translator.}
Rather than conditioning the denoiser at all timesteps, as in diffusion-based I2I methods such as \cite{Saharia2022Palette}, DMT learns a translator at one intermediate step \citep{Xia2025DMT}. Concretely, the translator $f_\theta$ is trained to map the noisy source representation $x_{t^\star}$ to the corresponding noisy target representation $y_{t^\star}$ as shown in Figure~\ref{fig:dmt_model}.  Under the DMT derivation, this yields a simple regression objective up to scaling by the diffusion variance:
\begin{equation}
\mathcal{L}_{\mathrm{DMT}}(t^\star)=
\mathbb{E}\left[\frac{1}{2(1-\bar{\alpha}_{t^\star})}\,\left\|f_\theta(x_{t^\star})-y_{t^\star}\right\|_2^2\right],
\label{eq:dmt_loss}
\end{equation}
where $y_{t^\star}$ is obtained by applying the same forward diffusion to $y_0$. Intuitively, $f_\theta$ learns the cross-domain transformation in a regime where small-scale details have been partially suppressed by noise, while global morphology (e.g., coronal hole extent) remains informative.

\textbf{Reverse diffusion in the target domain.}
After translation, DMT constructs an intermediate target-domain latent $\tilde{y}_{t^\star}$ and then runs the reverse denoising process of a DDPM trained on the target domain to generate $\tilde{y}_0$:
\begin{equation}
p_\phi(\tilde{y}_{0:T}) = p(\tilde{y}_T)\prod_{t=1}^{T} p_\phi(\tilde{y}_{t-1}\mid \tilde{y}_t),
\end{equation}
with $p_\phi(\tilde{y}_{t-1}\mid \tilde{y}_t)$ parameterized by a U-Net denoiser \citep{Ho2020DDPM,Ronneberger2015UNet}. In our implementation, we initialize the reverse chain at $t^\star$ using the translated latent and generate $\tilde{y}_0$ by sampling only from $t^\star \rightarrow 0$, which substantially reduces inference cost relative to full-length diffusion sampling. When needed, we further accelerate sampling using implicit samplers such as DDIM \citep{Song2021DDIM}.

\textit{Choosing the translation timestep.}
The choice of $t^\star$ governs a trade-off between (i) how \emph{close} the two domains are at the noisy level (larger $t$ makes $x_t$ and $y_t$ more similar and easier to translate) and (ii) how much \emph{conditioning signal} remains from $x_0$ (larger $t$ also makes $x_t$ less informative about $x_0$). DMT proposes selecting $t^\star$ by precomputing a distance curve between $(x_t,y_t)$ and a signal-retention curve between $(x_0,x_t)$ and choosing their intersection under an image-similarity measure (e.g., SSIM) \citep{Xia2025DMT,Wang2004SSIM}. We follow this principle to set $t^\star$ for the \HeI{}$\rightarrow$AIA~193~\AA{} translation task.

Overall, DMT allows us to (1) train a strong unconditional (or class-agnostic) diffusion prior on the target domain (AIA~193~\AA) using standard DDPM training \citep{Ho2020DDPM,SohlDickstein2015DPM,Song2021SDE}, and (2) learn an efficient, task-specific translator that performs the cross-domain mapping in a single step, while the subsequent denoising recovers realistic target-domain appearance and structure.

\subsection{CH-aware Diffusion Model Translator}\label{subusec:CH-aware DMT}
To support CH-aware evaluation and to steer the translator toward improved fidelity in low-emission structures, we augment the base DMT framework with CH-aware supervision. Specifically, we (i) construct deterministic, disk-restricted CH masks from AIA~193~\AA{} images using a consistent thresholding procedure and (ii) incorporate CH-aware loss terms during training that upweight reconstruction errors inside CH regions and emphasize boundary localization. We refer to this variant as \emph{CH-aware DMT}. The resulting CH masks are used as large-scale proxies for CH morphology (not as definitive expert segmentations) and provide a consistent, reproducible basis for comparisons between ground truth and synthetic outputs.

\subsubsection{CH Mask Construction (Ground Truth and Synthetic)}
To enable mask-based evaluation of CH morphology, we derive binary CH masks from both the \emph{ground-truth} and \emph{synthetic} AIA~193~\AA{} images through a disk-restricted intensity thresholding procedure. Each image is converted to grayscale and restricted to the on-disk region to exclude off-limb background. 
A per-image CH threshold $\tau$ is then computed using Otsu thresholding \citep{Otsu1979}. In our application, the raw Otsu threshold \(t_{\mathrm{Otsu}}\) separates darker low-emission regions from the brighter disk background. To better isolate the darkest CH regions, we apply a fixed negative offset of $-0.24$ in normalized intensity units, i.e., $\tau = t_{\mathrm{Otsu}} - 0.24$. Pixels satisfying $I < \tau$ are labeled as CH candidates. Small connected components with an area below 200 pixels are removed to suppress isolated noise and very small dark features. This procedure yields masks that serve as consistent, large-scale proxies for CH morphology rather than definitive segmentation of all CH structure, and it allows like-for-like mask comparisons between ground truth and synthetic images.

Formally, for the ground-truth AIA~193~\AA{} image $I_{\mathrm{GT}}$, the GT CH mask is
\begin{equation}
M_{\mathrm{GT}}(p)=\mathbb{I}\big(I_{\mathrm{GT}}(p) < \tau\big),
\end{equation}
where $p$ indexes disk pixels and $M_{\mathrm{GT}}\in\{0,1\}^{H\times W}$, and $\mathbb{I}(\cdot)$ denotes the indicator function, which equals 1 when its argument is true and 0 otherwise. 

For boundary-aware supervision and diagnostics, we compute an edge map from the GT mask using a morphological gradient \citep{Serra1982}:
\begin{equation}
E_{\mathrm{GT}} = \mathrm{dilate}(M_{\mathrm{GT}}) - \mathrm{erode}(M_{\mathrm{GT}}),
\end{equation}
implemented efficiently with local pooling operators. This concentrates boundary constraints on transition regions that define CH extent and shape. Here, \(\mathrm{dilate}(\cdot)\) expands CH regions outward by one local neighborhood, whereas \(\mathrm{erode}(\cdot)\) shrinks them inward by removing boundary pixels. Their difference isolates the CH boundary band used for edge-aware supervision.

At test time, we apply the \emph{same} normalization, grayscale conversion, disk restriction, and threshold rule to the \emph{synthetic} AIA~193~\AA{} output $\tilde{I}$ to obtain
\begin{equation}
M_{\mathrm{SYN}}(p)=\mathbb{I}\big(\tilde{I}(p) < \tau_{\mathrm{SYN}}\big),
\end{equation}
so that GT and synthetic masks are directly comparable by construction. In practice, we use the same threshold rule $\tau_{\mathrm{SYN}} = t_{\mathrm{Otsu}}(\tilde{I}) - 0.24$ applied to the synthetic image, mirroring the GT procedure.

\subsubsection{CH-aware Modifications to the DMT Training Objective}
While the base DMT objective promotes overall photometric consistency, CH-aware applications place disproportionate importance on (i) intensity fidelity inside CHs, (ii) sharp and correctly localized CH boundaries, and (iii) correct global CH extent under strong class imbalance (CH pixels are typically a minority). We therefore augment the generator-side objective with three CH-informed terms (in addition to the base reconstruction term), each targeting a distinct failure mode. Collectively, these terms encourage the model to allocate more capacity to CH-relevant regions and to preserve boundary placement more faithfully, motivating the name \emph{CH-aware DMT}.

\paragraph{(1) CH-weighted L1 content loss.}
We upweight pixel-wise reconstruction errors inside GT CH regions:
\begin{equation}
\mathcal{L}_{\mathrm{CH\text{-}L1}}
=\frac{1}{| \Omega |}\sum_{p\in\Omega}\,w(p)\,\big|\tilde{I}(p)-I_{\mathrm{GT}}(p)\big|,\quad
w(p)=1+(\lambda_{\mathrm{CH}}-1)\,M_{\mathrm{GT}}(p),
\end{equation}
where $\Omega$ denotes the disk pixels used for training and  $\lambda_{\mathrm{CH}}=5$ increases the penalty within CHs. This term explicitly prioritizes fidelity in low-intensity CH regions.

\paragraph{(2) Edge-aware boundary consistency loss.}
To discourage blurry transitions and boundary drift, we enforce gradient consistency between synthetic and GT images \emph{at CH edges}:
\begin{equation}
\mathcal{L}_{\mathrm{bdy}}
=\frac{\sum_{p\in\Omega} E_{\mathrm{GT}}(p)\,\big\|\nabla \tilde{I}(p)-\nabla I_{\mathrm{GT}}(p)\big\|_2^2}
{\sum_{p\in\Omega}E_{\mathrm{GT}}(p)+\epsilon}.
\end{equation}
Restricting the penalty to $E_{\mathrm{GT}}$ targets contour localization rather than general texture matching. Here, $\epsilon=10^{-6}$ is a small numerical-stability constant that prevents division by zero in the rare case that no CH-edge pixels are present in the mask. 

\paragraph{(3) Dice segmentation loss with an auxiliary CH head.}
We attach a lightweight segmentation head $s(\cdot)$ to the synthetic output and train it to predict the GT CH mask. With $P=\sigma(s(\tilde{I}))\in[0,1]^{H\times W}$, we use a Dice-style loss robust to class imbalance \citep{Dice1945,Sorensen1948,Milletari2016,Sudre2017}:
\begin{equation}
\mathcal{L}_{\mathrm{Dice}}
=1-\frac{2\sum_{p\in\Omega} P(p)\,M_{\mathrm{GT}}(p)+\epsilon}
{\sum_{p\in\Omega} P(p)+\sum_{p\in\Omega} M_{\mathrm{GT}}(p)+\epsilon}.
\end{equation}
This term adds a global, shape-level constraint that complements the local photometric and boundary objectives. 

\paragraph{Combined objective and hyperparameters.}
The final generator objective is the sum of the base reconstruction loss, $\mathcal{L}_{\mathrm{base}}$, and the three CH-aware terms:
\begin{equation}
\mathcal{L}_{\mathrm{total}}
=\mathcal{L}_{\mathrm{base}}
+\mathcal{L}_{\mathrm{CH\text{-}L1}}
+\lambda_{\mathrm{bdy}}\,\mathcal{L}_{\mathrm{bdy}}
+\lambda_{\mathrm{seg}}\,\mathcal{L}_{\mathrm{Dice}}. 
\end{equation}
In all experiments, we set the CH-aware loss weights to $\lambda_{\mathrm{bdy}}=2$ and $\lambda_{\mathrm{seg}}=2$, keeping these values fixed for the reported results.
We expect these changes to improve CH-oriented fidelity because they (i) focus intensity accuracy inside CHs, (ii) enforce sharp, well-localized CH boundaries through edge-aware constraints, and (iii) regularize global CH extent via a Dice-based shape objective designed for imbalanced segmentation \citep{Milletari2016,Sudre2017}.

\section{Results}\label{sec:results}
Table~\ref{tab:results_summary} summarizes the datasets and evaluation framework used throughout Section~\ref{sec:results}. For each subsection, we present: 
1) the \HeI{} input source used to generate the synthetic AIA~193~\AA{} images; 
2) the time span analyzed; and 
3) the corresponding ground-truth or external comparison product used for quantitative evaluation or qualitative morphology comparison. 
This structure organizes the evidence according to the available validation source. 
Section~3.1 uses direct SDO/AIA~193~\AA{} ground truth for the main quantitative evaluation. 
Section~3.2 uses SOHO/EIT~195~\AA{} full-disk observations and SOHO/EIT~195~\AA{} synoptic maps for cross-instrument EUV comparisons in the pre-SDO interval. 
Section~3.3 uses Yohkoh/SXT soft X-ray observations for qualitative evaluation of large-scale coronal morphology during the KPVT era. 
Section~3.4 uses independent solar-activity proxies, including sunspot number and F10.7 radio flux, to assess long-term temporal consistency of the reconstructed disk-integrated signal.



\begin{table}[!htbp]
\centering
\caption{Summary of datasets, time ranges, and ground-truth comparison products used in Section~\ref{sec:results}.}
\label{tab:results_summary}
\begin{tabular}{llll}
\toprule
\textbf{Section} & \textbf{Input (source)} & \textbf{Time range} & \textbf{Ground-truth/Comparison} \\
\midrule
\ref{sec:res_31} &
NSO/SOLIS &
2011--2015 &
SDO/AIA~193~\AA{}\\
\ref{sec:res_eit195} &
NSO/SOLIS &
2005--2015 &
SOHO/EIT~195~\AA{}\\
\ref{sec:res_33} &
NSO/KPVT &
1974--1993 &
Yohkoh/SXT soft X-ray\\
\ref{sec:res_emission} &
NSO/KPVT; NSO/SOLIS&
1974--1993; 2005--2015 &
F10.7 and sunspots number \\
\bottomrule
\end{tabular}
\end{table}

\begin{figure*}[!htbp]
    \centering
    \includegraphics[width=\textwidth]{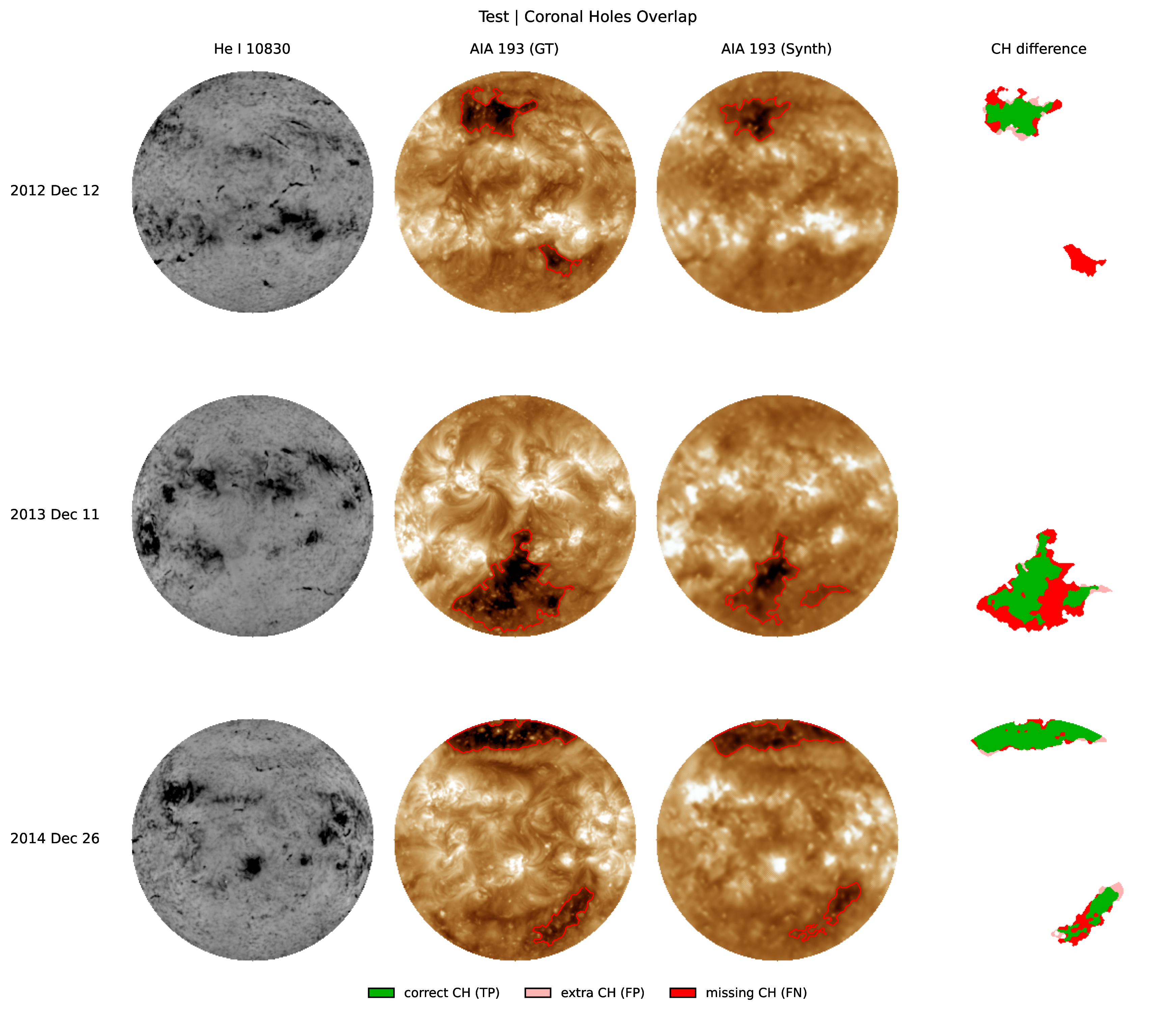}
    \caption{\textbf{CH reconstruction on the AIA~193~\AA{} test set (2011--2015).}
    For three representative test dates, we show: \emph{(left)} the conditioning \HeI{} input; \emph{(middle-left)} the GT AIA~193~\AA{} image; \emph{(middle-right)} the synthetic AIA~193~\AA{} image generated by CH-aware DMT; and \emph{(right)} the CH mask accuracy map.
    CH masks are derived independently from GT and synthetic AIA~193~\AA{} using Otsu thresholding \citep{Otsu1979}.
    Green indicates correctly recovered CH pixels (TP), pink indicates extra CH pixels in the synthetic image (FP), and red indicates missing CH pixels relative to the GT mask (FN).}
    \label{fig:ch_test_aia}
\end{figure*}
\subsection{Test-Set Performance on SDO/AIA~193~\AA{} from 2011 to 2015}\label{sec:res_31} 

We first evaluate the CH-aware DMT model on the held-out SDO/AIA~193~\AA{} test set by comparing (i) reconstructed EUV morphology in the synthetic AIA~193~\AA{} images and (ii) accuracy CH masks derived from the ground-truth (GT) and synthetic images. CH masks are extracted using Otsu thresholding \citep{Otsu1979} on the AIA~193~\AA{} intensity distribution, and the same segmentation procedure is applied to synthetic outputs to enable a direct comparison.

Figure~\ref{fig:ch_test_aia} shows three representative dates from the test set, displaying the input \HeI{} spectroheliogram, the GT AIA~193~\AA{} image, the synthetic AIA~193~\AA{} reconstruction, and a CH mask difference map decomposed into true positives (TP), false positives (FP), and false negatives (FN). Across these examples, the model reconstructs the dominant full-disk AIA~193~\AA{} morphology and recovers CH-related low-intensity regions with strong overlap (TP regions) under the chosen masking procedure. Residual discrepancies in CH \emph{extent} (FP/FN regions) are expected due to the different formation physics of \HeI{}, where chromospheric absorption modulated by coronal irradiance and atmospheric coupling, compared with the coronal iron emission that dominates the AIA~193~\AA{} band. Table~\ref{tab:metrics_chaware_vs_dmt} reports PSNR, correlation coefficient (CC), and mean absolute error (MAE) for the CH-aware DMT for the test-set. All metrics are computed using RGB images normalized to the $[0,1]$ intensity range; thus, MAE represents the average absolute normalized intensity difference, PSNR is computed with a data range of 1.0, while CC is invariant to uniform intensity scaling. It shows high accuracy on full-disk morphology (CC=0.9227) with absolute error (MAE=0.0711). Performance degrades in the CH-restricted evaluation region, as expected for the lowest-intensity structures (CC=0.8408; MAE=0.1086), but remains consistent to preserve large-scale low-emission morphology under a masking procedure.

\begin{table}[!htbp]
\centering
\caption{Quantitative evaluation of the synthetic AIA193\AA{} images generated by the CH-aware DMT model on the 2011–2015 test set (December). Metrics are reported for both full-disk images and coronal-hole regions. Higher values indicate better performance for PSNR and CC, while lower values indicate better performance for MAE.}
\label{tab:metrics_chaware_vs_dmt}
\begin{tabular}{lrrr}
\toprule
Region & PSNR $\uparrow$ & CC $\uparrow$ & MAE $\downarrow$\\
\midrule
Full disk      & 20.3081 & 0.9227 & 0.0711 \\
Coronal holes  & 17.4416 & 0.8408 & 0.1086 \\
\bottomrule
\end{tabular}
\end{table}
\begin{figure*}[!htbp]
    \centering
    \includegraphics[width=\textwidth]{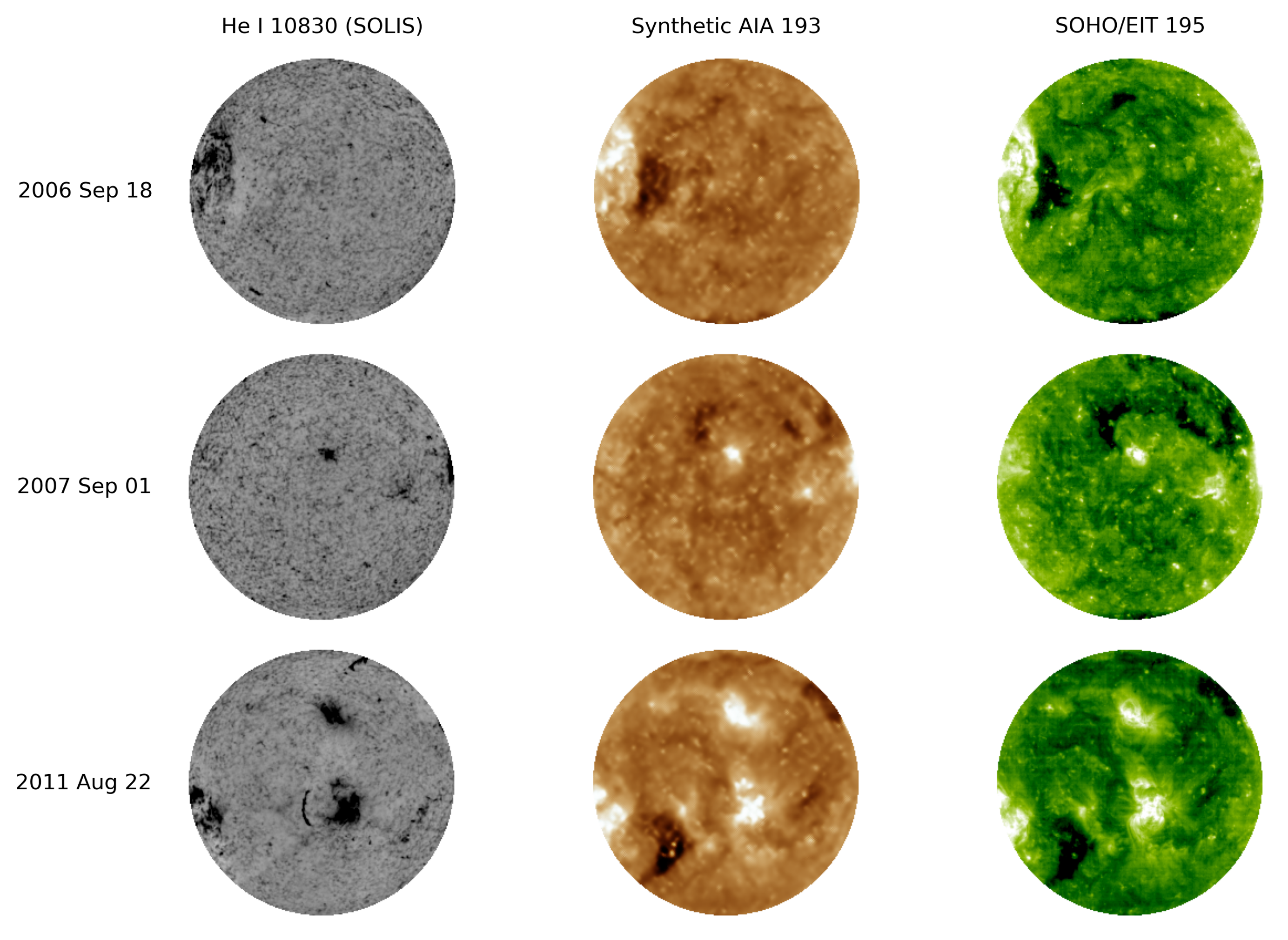}
    \caption{\textbf{Evaluation over 2005--2015 using SOHO/EIT~195~\AA.}
    For three representative dates, we show (left-to-right) the SOLIS \HeI{} input, the CH-aware DMT-generated synthetic AIA~193~\AA{} reconstruction, and the contemporaneous SOHO/EIT~195~\AA{} full-disk image.}
    \label{fig:val_eit_2005_2015}
\end{figure*}

\subsection{Cross-Instrument Evaluation with SOHO/EIT~195~\AA{} from 2005 to 2015}\label{sec:res_eit195}
To assess whether the learned mapping from \HeI{} to EUV image remains physically plausible outside the direct SDO/AIA training regime, we compare reconstructions against SOHO/EIT~195~\AA{} observations from 2005 to 2015, an interval covered by SOLIS \HeI{} and by SOHO/EIT. Because EIT~195~\AA{} and AIA~193~\AA{} have closely related temperature sensitivity dominated by Fe~\textsc{xii} under non-flaring conditions, EIT provides a natural external EUV reference for evaluating large-scale coronal morphology. In this subsection, we present two complementary cross-instrument checks: (i) full-disk, single-date comparisons against contemporaneous SOHO/EIT~195~\AA{} images, and (ii) Carrington-rotation synoptic-map comparisons using the EIT~195~\AA{} synoptic intensity archive from Oulu. Throughout, we emphasize qualitative accuracy in global structure and contrast ordering rather than pixel-level correspondence, given differences in instrument response, cadence, and synoptic-map construction.

\subsubsection{Full-Disk Comparison Against SOHO/EIT~195~\AA{}}\label{sec:res_eit195_fulldisk}
To test generalization beyond the direct SDO/AIA overlap and to assess physical plausibility with an independent coronal diagnostic, we compare synthetic AIA~193~\AA{} images against SOHO/EIT~195~\AA{} observations. EIT~195~\AA{} is an appropriate EUV benchmark over 2005--2015 because it samples coronal Fe~\textsc{xii} emission near $\sim$1--1.5~MK, broadly similar to the primary contribution in AIA~193~\AA{} under non-flaring conditions \citep{Delaboudiniere1995,ODwyer2010,Lemen2012,Boerner2012}. Although passbands, cadence, and spatial response differ between instruments, EIT~195~\AA{} and AIA~193~\AA{} are commonly treated as closely related EUV coronal proxies for synoptic studies, particularly after appropriate preprocessing/normalization in multi-instrument mapping pipelines \citep{Caplan2016,Hamada2020}. Accordingly, we interpret this comparison at the level of large-scale morphology and contrast ordering rather than pixel-level accuracy. Figure~\ref{fig:val_eit_2005_2015} presents three illustrative dates spanning 2005--2015, showing the SOLIS \HeI{} input, the resulting synthetic AIA~193~\AA{} reconstruction, and the corresponding SOHO/EIT~195~\AA{} full-disk image. The synthetic reconstructions reproduce the dominant large-scale coronal morphology visible in EIT, including extended low-emission regions and bright active-region complexes, supporting qualitative cross-instrument consistency of the reconstructed global structure.

\begin{figure*}[!htbp]
    \centering
    \includegraphics[width=\textwidth]{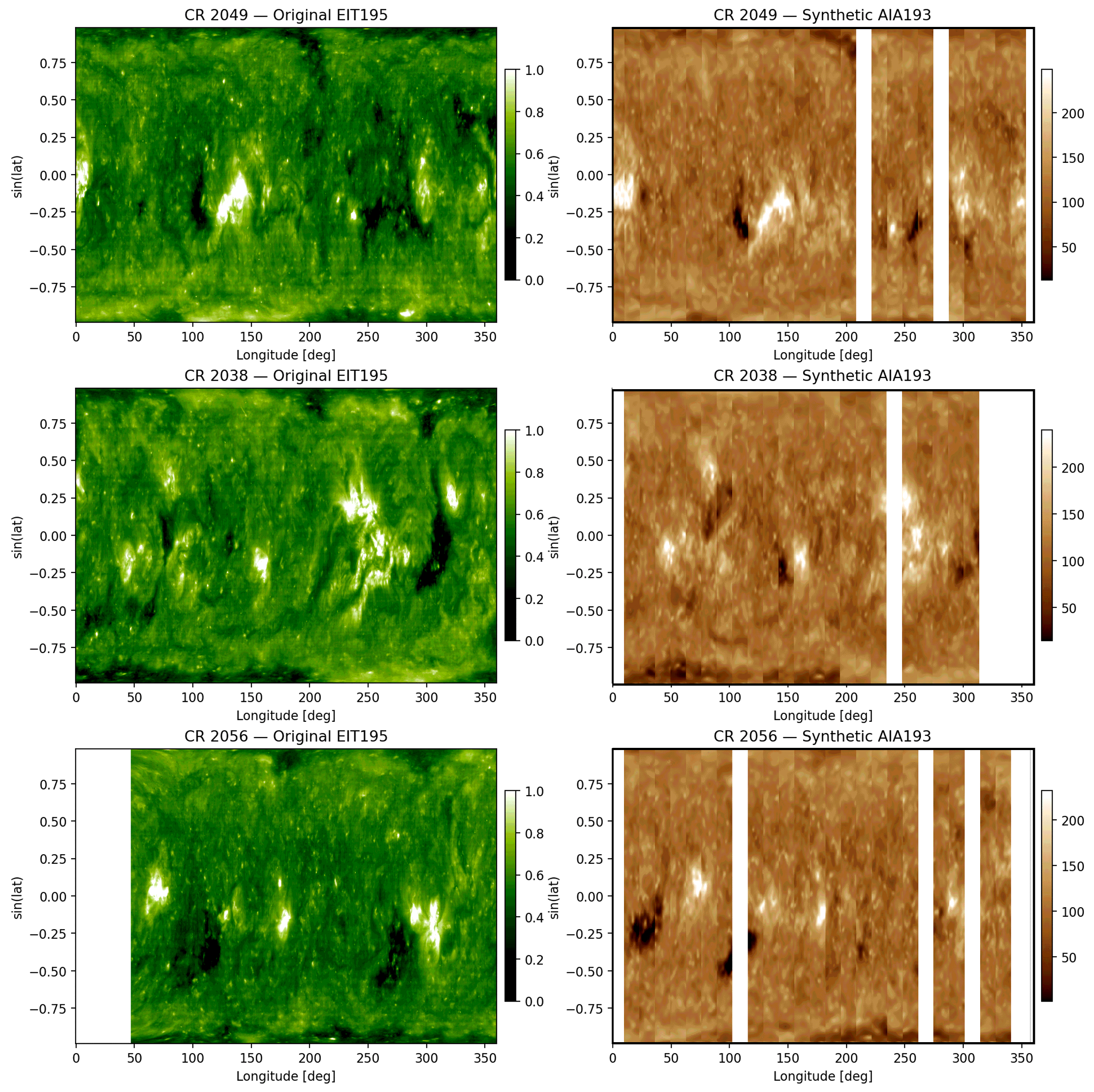}
    \caption{\textbf{Qualitative comparison of Carrington synoptic intensity maps of ground truth EIT~195~\AA{} and synthetic AIA~193~\AA{} reconstructions.}
    Each row shows a representative Carrington rotation. The left column displays the gt EIT~195~\AA{} synoptic intensity map, while the right column shows the corresponding Carrington synoptic map constructed from the synthetic AIA~193~\AA{} full-disk sequence.}
    \label{fig:carrington_map_comparison}
\end{figure*}
\subsubsection{Carrington-Rotation Synoptic-Map Comparison Using SOHO/EIT~195~\AA{} Archive}\label{sec:res_eit195_synoptic}
In addition to image-level validation using full-disk EIT 195 images, we construct Carrington-rotation maps to evaluate whether the large-scale coronal structures are reproduced coherently over time. Figure~\ref{fig:carrington_map_comparison} compares Carrington synoptic intensity maps from the SOHO/EIT~195~\AA{} archive with synoptic maps constructed from the synthetic AIA~193~\AA{} sequence for full rotations.
Across the selected rotations, the synthetic maps preserve the corona features.
Importantly, the main CH regions are spatially aligned with the corresponding low-emission structures in the EIT~195~\AA{} reference maps, indicating that the model recovers the broad rotation-scale morphology rather than only producing plausible full-disk images.
Because the synthetic maps are built from daily central-meridian strips, gaps and vertical discontinuities reflect incomplete temporal sampling rather than physical emission structure.
Therefore, this comparison is interpreted qualitatively, emphasizing large-scale CH placement and coronal organization rather than pixel-level accuracy.

\begin{figure*}[!htbp]\vspace{-0.2cm}
    \centering
    \includegraphics[width=\textwidth]{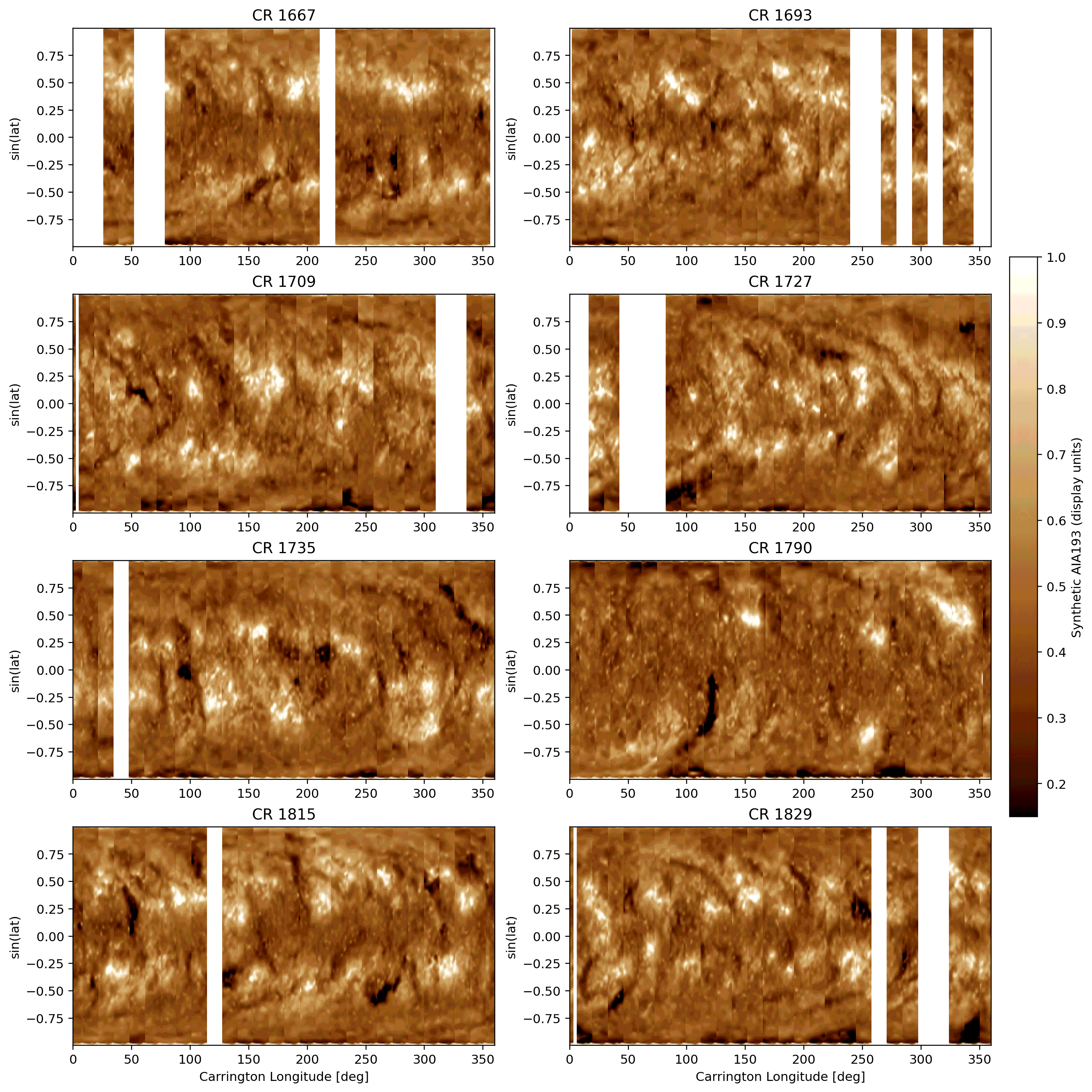}
    \caption{\textbf{Synthetic AIA~193~\AA{} Carrington synoptic maps reconstructed from KPVT \HeI{} observations.}
    Each panel shows a Carrington synoptic map constructed from the synthetic AIA~193~\AA{} full-disk sequence generated by applying the trained translator to KPVT \HeI{} inputs from 1974 to 1993. The maps are assembled at a 1-day cadence using central-meridian strips, so gaps indicate undersampled longitudes for a given rotation.}
    \label{fig:carrington_map_kpvt_synth}
\end{figure*}

\subsection{Evaluation of Synthetic AIA~193~\AA{} reconstructed from KPVT \HeI{} from 1974 to 1993}\label{sec:res_33}

We next assess historical applicability by applying the trained translator to the NSO/KPVT \HeI{} archive from 1974 to 1993 to generate a long-baseline synthetic AIA~193~\AA{} sequence. Because no contemporaneous full-disk EUV ground truth exists for most of this interval, we focus on two complementary, physics-motivated checks: (i) Carrington-rotation synoptic maps constructed from the synthetic sequence to visualize global organization over full rotations, and (ii) a morphology comparison against Yohkoh/SXT soft X-ray observations during the 1992 overlap period, where an independent coronal diagnostic is available.

\subsubsection{Carrington-Rotation Synoptic Maps Reconstructions}\label{sec:res_kpvt_carrington_map}
To show the long-term coronal image reconstruction produced by this framework, Figure~\ref{fig:carrington_map_kpvt_synth} shows synthetic AIA~193~\AA{} Carrington synoptic maps constructed from KPVT \HeI{} inputs between 1974 and 1993. No contemporaneous EUV synoptic maps are available for this interval, so direct validation is not possible. However, the maps still provide a useful qualitative view of the inferred global coronal structure. They are assembled at a 1-day cadence using central-meridian strips; therefore, the longitudinal gaps indicate missing observing days rather than the physical absence of emission.

The selected rotations sample different phases of Solar Cycles~21 and~22. CR~1667, around April--May 1978, falls in the rising phase of Solar Cycle~21, while CR~1693, CR~1709, CR~1727, and CR~1735, from 1980 to 1983, cover the maximum and early declining phase of the same cycle. The larger number of bright active-region structures in these maps is consistent with the higher level of activity expected near solar maximum. In contrast, CR~1790, around June--July 1987, is close to the minimum between Solar Cycles~21 and~22 and shows fewer strong active regions. CR~1815 and CR~1829, around April--May 1989 and May--June 1990, correspond to the rising phase and maximum of Solar Cycle~22, where active-region emission increases again. The latitude of the bright activity bands is also broadly consistent with the solar butterfly pattern: the bands migrate toward lower latitudes as Solar Cycle~21 evolves, while after the minimum, new-cycle activity reappears at higher latitudes before moving equatorward. Therefore, the amount and latitude of active-region emission seen in the synthetic maps are consistent with the expected solar-cycle phase, although this comparison remains qualitative.


\begin{figure*}[!htbp]
    \centering
    \includegraphics[width=\textwidth]{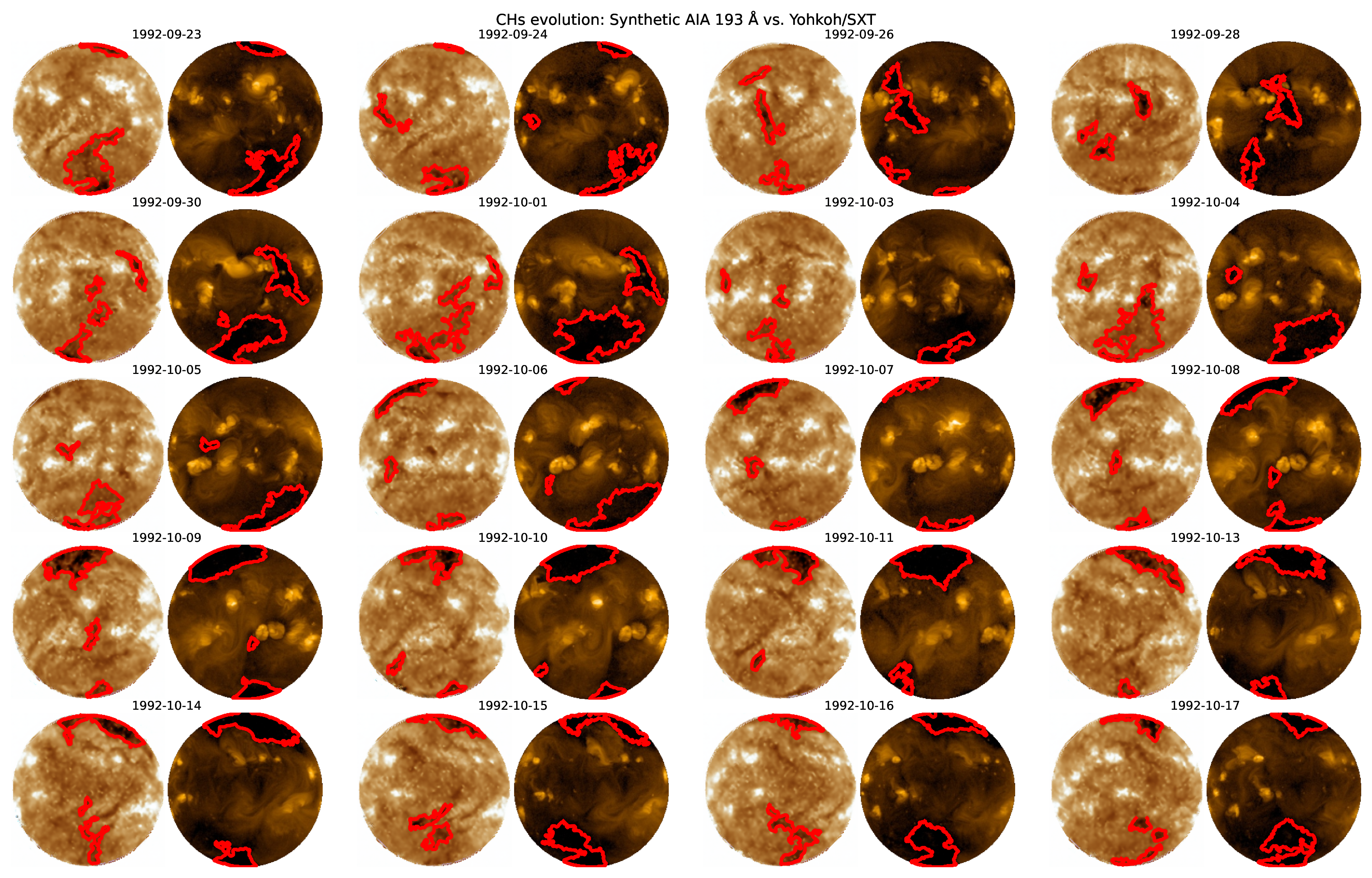}
    \caption{\textbf{Morphological comparison of synthetic AIA~193~\AA{} derived from KPVT and Yohkoh/SXT soft X-ray.}
    Sequence from 1992-09-23 to 1992-10-17 ($n=20$), showing paired synthetic AIA~193~\AA{} reconstructions generated from KPVT \HeI{} and contemporaneous Yohkoh/SXT images.}
    \label{fig:kpvt_vs_sxt}
\end{figure*}

\subsubsection{Morphology Comparison Using Yohkoh/SXT}\label{sec:res_yohkoh}
To obtain an external coronal reference within the KPVT interval, we leverage Yohkoh soft X-ray observations from the Soft X-ray Telescope (SXT), which provides full-disk coronal imaging in the early 1990s \citep{Tsuneta1991,Acton2016}. Although SXT has a different temperature response than the AIA~193~\AA{} passband, low-emission regions manifest in both EUV and soft X-ray diagnostics due to reduced emissivity in lower-density coronal structures \citep{Kahler1983,Cranmer2009}. We therefore use SXT as a morphological reference to evaluate whether low-emission structures in the synthetic AIA~193~\AA{} sequence show temporal evolution consistent with SXT observations.

Figure~\ref{fig:kpvt_vs_sxt} compares a time sequence of synthetic AIA~193~\AA{} reconstructions generated from KPVT \HeI{} inputs with temporally aligned Yohkoh/SXT observations from 1992-09-23 to 1992-10-17 ($n=20$). The paired views show coherent evolution of low-emission regions across the two diagnostics, supporting the physical plausibility of the reconstructed large-scale coronal structure during this overlap period. We emphasize that this comparison is qualitative and morphology-based, given differences in instrument response, contrast, and temperature sensitivity between soft X-ray and EUV diagnostics.

\begin{figure*}[!htbp]
    \centering
    \includegraphics[width=\textwidth]{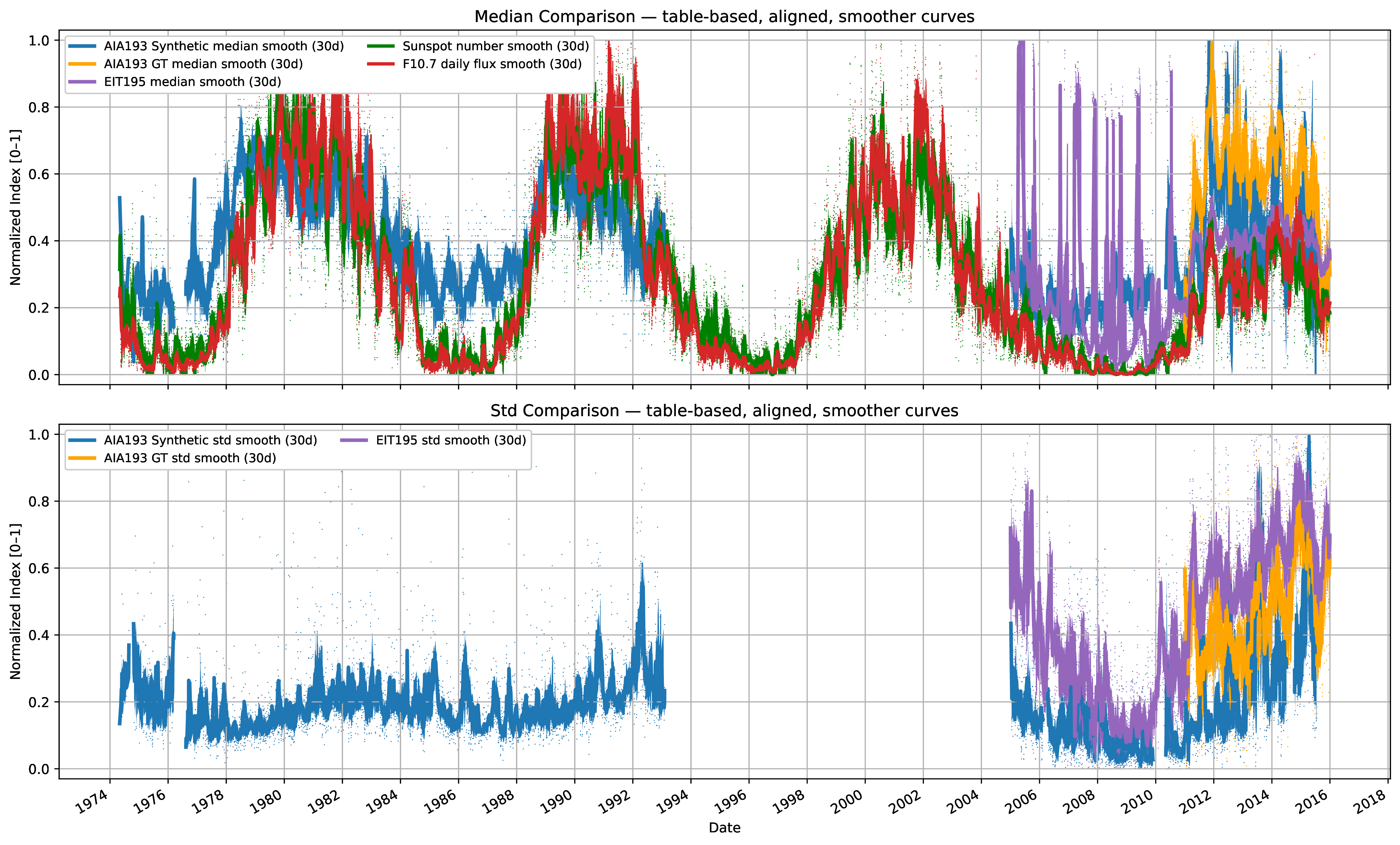}
    \caption{\textbf{Long-term normalized comparison of disk-integrated EUV statistics and solar-activity proxies.}
    Daily full-disk median and standard deviation statistics are computed for the synthetic AIA~193~\AA{} sequence and, where available, for observational GT AIA~193~\AA{} and SOHO/EIT~195~\AA{} series. Each time series is aligned in time, independently normalized to [0,1], and smoothed with a 30-day moving average. \emph{Top:} normalized median comparison including synthetic AIA~193~\AA{}, GT AIA~193~\AA{}, EIT~195~\AA{}, sunspot number, and F10.7 radio flux. \emph{Bottom:} normalized standard deviation comparison for synthetic AIA~193~\AA{}, GT AIA~193~\AA{}, and EIT~195~\AA{}.}
    \label{fig:emission_timeseries}
\end{figure*}

\subsection{Long-Term Normalized Trend Consistency and Comparison with Independent Solar-Activity Proxies}\label{sec:res_emission}

We next assess whether the reconstructed EUV signal preserves physically meaningful long-term temporal behavior by comparing disk-integrated summary statistics from the synthetic AIA~193~\AA{} sequence against independent observational EUV series and solar-activity proxies. Figure~\ref{fig:emission_timeseries} shows normalized time series of the full-disk median and standard deviation derived from the synthetic AIA~193~\AA{} product, together with corresponding curves from ground-truth AIA~193~\AA{} and SOHO/EIT~195~\AA{} where available. In the median panel, we additionally include the international sunspot number and the F10.7 radio flux (NOAA SWPC 10.7) as external activity references. We use the daily total sunspot number from the SILSO Version 2.0 record as the sunspot proxy \citep{SILSO_Sunspot_Number}, and the daily 10.7\,cm (2800\,MHz) solar radio flux record as the F10.7 proxy \citep{NOAA_SWPC_F107}. Because these series come from heterogeneous instruments and products with different radiometric scales, observing cadences, and long-term stability characteristics, the comparison is performed in terms of normalized indices rather than absolute intensities.

For the synthetic series, the 2005--2015 interval is generated from SOLIS \HeI{} observations, while the 1974--1993 interval is generated from KPVT \HeI{} observations; both are processed through the same trained translator to form a long-baseline synthetic AIA~193~\AA{} record. For each dataset, we compute daily full-disk summary statistics after masking invalid off-disk pixels. The resulting daily series are aligned in time, independently normalized to the interval [0,1], and smoothed with a 30-day moving average to suppress short-timescale fluctuations while retaining broad solar-cycle-scale modulation.

The synthetic AIA~193~\AA{} time series captures the expected solar-cycle modulation of coronal emission. The mean decreases during low-activity intervals and increases during more active phases, reflecting the reduced EUV brightness and weaker active-region contrast near solar minimum and the enhanced coronal emission during active periods. These long-term variations track the sunspot number and the F10.7~cm radio flux, suggesting that the reconstruction retains cycle-scale information encoded in the \HeI{} observations, rather than only reproducing plausible instantaneous morphology.

\begin{table*}[!htbp]
\centering
\caption{
Monthly time-series accuracy between the reconstructed synthetic AIA~193~\AA{} signal and available reference series.
Daily values are first averaged into monthly means, so each month contributes one data point.
For each comparison, the two overlapping monthly series are restricted to the same dates and normalized to $[0,1]$ before computing the metrics.}
\label{tab:timeseries_metrics_by_period}
\begin{tabular}{lllrrr}
\hline
Period & Statistic & Reference & CC & MAE & RMSE \\
\hline
\multirow{2}{*}{1974--1993} 
& Median & Sunspot number & 0.843 & 0.452 & 0.554 \\
& Median & F10.7 radio flux & 0.830 & 0.603 & 0.649 \\
\hline
\multirow{4}{*}{2005--2015} 
& Median & SOHO/EIT 195~\AA{} & 0.228 & 0.449 & 0.491 \\
& Median & Sunspot number & 0.320 & 0.382 & 0.496 \\
& Median & F10.7 radio flux & 0.527 & 0.635 & 0.660 \\
& Std. dev. & SOHO/EIT 195~\AA{} & 0.674 & 0.555 & 0.569 \\
\hline
\multirow{2}{*}{2011--2015} 
& Median & GT AIA 193~\AA{} & 0.479 & 0.505 & 0.521 \\
& Std. dev. & GT AIA 193~\AA{} & 0.545 & 0.628 & 0.642 \\
\hline
\end{tabular}
\end{table*}

Table~\ref{tab:timeseries_metrics_by_period} reports the monthly time-series accuracy between the reconstructed synthetic AIA193\AA{} signal and the available reference series shown in Figure~\ref{fig:emission_timeseries}. 
Daily disk-level statistics were first averaged within each calendar month, so that each month contributes one data point. 
For each comparison, the synthetic and reference monthly series were restricted to their overlapping dates, normalized to $[0,1]$, and compared using CC, MAE, and RMSE. 
Therefore, these metrics evaluate monthly temporal variability rather than pixel-level image accuracy.

During the KPVT period, 1974--1993, the synthetic median signal shows strong accuracy with independent solar-activity proxies, with CC values of 0.843 against sunspot number and 0.830 against F10.7 radio flux. 
This indicates that the reconstructed median coronal-emission signal broadly follows the long-term solar-cycle modulation. 

For the SOLIS period, 2005--2015, the median accuracy is weaker, with CC values of 0.228 against SOHO/EIT195\AA{}, 0.320 against sunspot number, and 0.527 against F10.7 radio flux. 
The lower accuracy with sunspot number and F10.7 is expected because these indices have a different physical nature from AIA193\AA{} emission;
therefore, they are mainly useful for qualitative evaluation of solar-cycle phase consistency rather than direct month-by-month intensity validation.  The lower median accuracy with EIT195\AA{} may also be affected by known calibration and degradation-related temporal drifts in long-term EIT/AIA EUV intensities~\citep{Hamada2020}. 
However, the stronger standard deviation accuracy with EIT195\AA{} (CC=0.674) suggests that the temporal evolution of disk-level coronal-emission contrast is more consistent with the synthetic reconstruction than the absolute median-emission trend.

During the direct SOLIS/AIA overlap period, 2011--2015, the synthetic sequence shows moderate accuracy with GT AIA193\AA{}, with CC values of 0.479 for the median and 0.545 for the standard deviation. It is worth noting that part of the discrepancy in the accuracy metrics may be related to the different data formats used to construct the reference and synthetic time series. The GT AIA~193~\AA{} curve was computed from calibrated FITS files, whereas the synthetic curve was computed from PNG outputs. This difference in data format and intensity scaling can affect the normalized full-disk emission values, leading to differences in the resulting time-series metrics. Because GT AIA193\AA{} is the closest available reference to the reconstructed quantity, this provides the most direct validation of the monthly reconstructed signal. 

Overall, the monthly metrics indicate that the reconstructed sequence preserves broad solar-cycle variability during the historical KPVT interval and captures meaningful disk-level coronal-structure variability during the SOLIS/AIA overlap period. 
The relatively low accuracy in some rows should therefore be interpreted in light of differences in physical origin, passband, calibration, and instrument degradation, rather than as a direct failure of the reconstruction.

\section{Discussion}\label{sec:discussion_old}

\subsection{Robustness and Limitations of the CH-aware DMT}
The goal of this work is to extend EUV coronal context backward in time using the long synoptic record of \HeI{} observations, instead of reproducing absolutely calibrated AIA~193~\AA{} radiometry. Because \HeI{} absorption is modulated by coronal irradiance and magnetic topology, it contains indirect information about the overlying corona. The CH-aware DMT exploits this relationship to reconstruct a structural coronal proxy. 

On the direct AIA test set, the reconstructions preserve full-disk morphology and recover major low-emission regions. Over longer time intervals, disk-integrated statistics follow broad solar-cycle modulation consistent with independent activity proxies such as sunspot number and F10.7. At the same time, synoptic-map comparisons show coherent longitudinal and latitudinal organization of extended low-emission regions and bright active-region complexes. Thus, the model reproduces the large-scale coronal structure and its temporal evolution.

The model also has limitations.  Absolute intensity calibration is not preserved, and direct photometric comparison is complicated by differences among calibrated FITS products, rendered image formats, and cross-instrument intensity scales. Similarly, CH-derived products should be interpreted as approximate and method-dependent. Low-intensity regions may correspond to open-field coronal holes, but without magnetic field information, it is difficult to determine whether unipolar regions dominate them. Therefore, CH masks and synoptic summaries are used here to demonstrate the types of analyses enabled by the reconstructed long-baseline AIA~193~\AA{} product, rather than as definitive measurements of CH boundaries or areas.


\subsection{Cross-instrument Validation with Independent Coronal and Activity Proxies}
A central question is whether CH-aware DMT preserves coronal structure beyond the specific SDO/AIA training domain. SOHO/EIT~195~\AA{} provides the closest pre-SDO EUV reference because EIT~195~\AA{} and AIA~193~\AA{} are both dominated by Fe~\textsc{xii} emission under non-flaring conditions \citep{Delaboudiniere1995,ODwyer2010,Lemen2012,Boerner2012}. Accordingly, we compare the synthetic AIA~193~\AA{} sequence with SOHO/EIT~195~\AA{} at two levels: full-disk morphology on matched dates and Carrington-rotation synoptic structure using SOHO/EIT~195~\AA{} synoptic-map. These comparisons show that the reconstructions reproduce the low-emission regions, bright active-region complexes, and broad longitudinal and latitudinal organization seen in products reconstructed from EIT. Because the instruments and synoptic-map pipelines differ in response, preprocessing, cadence, and normalization, this accuracy is interpreted qualitatively rather than as pixel-level validation \citep{Caplan2016,Hamada2020}.

For earlier epochs without routine full-disk EUV observations, Yohkoh/SXT provides an additional coronal reference. Although soft X-ray emission has a different temperature response than EUV 193/195~\AA{}, extended low-emission coronal structures can appear in both diagnostics because of reduced density and emissivity \citep{Kahler1983,Cranmer2009}. The qualitative accuracy between synthetic AIA~193~\AA{} reconstructed from KPVT reconstructions and Yohkoh/SXT images during the 1992 overlap period support the physical plausibility of the reconstructed large-scale corona outside the training interval.

Finally, the disk-integrated synthetic AIA~193~\AA{} signal follows broad cycle-scale variations consistent with GT AIA~193~\AA{}, SOHO/EIT~195~\AA{}, sunspot number, and F10.7 radio flux. While the EIT record can include additional fluctuations related to instrumental degradation and long-term stability effects \citep{Hamada2020}, agreement with both EUV observations and independent activity proxies indicates that the translator preserves meaningful temporal information, not only plausible instantaneous morphology.

\subsection{Implications for Solar-cycle Studies and Historical Reconstructions}
CH-aware DMT provides a practical pathway to extend AIA~193~\AA{} coronal context across multiple solar cycles using chromospheric \HeI{} archives. The most immediate value of the approach is to deliver a long-baseline and reliable proxy for EUV coronal structure that supports historical analyses of coronal evolution over time intervals that previously lacked routine EUV imaging. Such a proxy can enable studies of long-term changes in global coronal morphology, cycle-to-cycle variability, hemispheric asymmetry, and the evolution of low-emission regions associated with open magnetic flux.

CH-aware analyses represent one promising downstream application, but they should be pursued with appropriate caution and validation. In particular, the reconstructed EUV products can be used to generate candidate CH maps 
or establish CH archives under controlled protocols. More broadly, this framework opens a route to building multi-decade coronal proxy products from pre-SDO and pre-SOHO chromospheric records, supporting future work on open-flux evolution and its coupling to heliospheric structure.

\section{Conclusions}\label{sec:conclusion}

We presented a diffusion-based conditional translation approach, CH-aware Diffusion Model Translator, to reconstruct synthetic AIA~193~\AA{} images from \HeI{} observations with the primary objective of extending EUV coronal context backward in time. Applying the trained model to the KPVT \HeI{} archive from 1974 to 1993 and the SOLIS \HeI{} archive from 2005 to 2015 produces a long-baseline synthetic AIA~193~\AA{} sequence that connects pre-SDO and SDO-era intervals. Rather than targeting absolute radiometric fidelity, this work focuses on reconstructing physically plausible large-scale coronal structure and its temporal evolution over decadal baselines. Our main conclusions are:

\begin{itemize}
\item \textbf{The reconstructed sequence preserves long-term temporal behavior in total coronal emission.}
Disk-integrated statistics show that the synthetic AIA~193~\AA{} sequence captures broad cycle-scale modulation over the KPVT and SOLIS intervals. In particular, the temporal evolution of the reconstructed total emission is consistent with independent solar-activity proxies, such as sunspot number and F10.7 radio flux, supporting its use as a long-term coronal-emission proxy. The accuracy is not expected to be exact because these proxies measure different physical quantities than EUV disk emission, but their consistency provides evidence that the reconstruction preserves meaningful temporal variability.

\item \textbf{The reconstructions preserve coherent large-scale spatial morphology.}
Qualitative comparisons using full-disk images and Carrington synoptic maps show that the synthetic AIA~193~\AA{} images retain physically plausible global coronal organization, including active-region bands, coronal-hole-like structures, and large-scale latitudinal patterns. The synoptic maps further show morphology that is consistent with the expected solar-cycle phase, including changes in active-region density and latitude over time. These results indicate that CH-aware DMT reconstructs spatially coherent coronal structure, rather than only reproducing disk-integrated trends.

\item \textbf{Cross-instrument comparisons support the physical plausibility of the reconstructions.}
Comparisons with SOHO/EIT~195~\AA{} synoptic products and, for earlier epochs, Yohkoh/SXT morphology provide external evidence that the reconstructions capture meaningful coronal patterns beyond the direct AIA training interval. These checks support the interpretation of CH-aware DMT outputs as structural and temporal EUV proxies, rather than as calibrated EUV measurements.
\end{itemize}

\noindent \textbf{Next steps.} Future work will (i) incorporate uncertainty estimation to flag ambiguous regimes and guide downstream use, (ii) develop improved intensity homogenization and cross-instrument normalization for more quantitative comparisons, (iii) refine synoptic-map construction and evaluation protocols, including cadence, strip width, gap handling, and consistency checks, to better support long-term morphological analyses, and (iv) expand validation using additional historical coronal references and targeted expert labeling to better quantify the reliability of derived CH products.


\section*{Acknowledgments}
This research is supported by NASA grants 80NSSC24K0548 and 80NSSC24M0174. This work made use of data from the Solar Dynamics Observatory/Atmospheric Imaging Assembly (SDO/AIA; \citealt{Lemen2012}), the Solar and Heliospheric Observatory/Extreme Ultraviolet Imaging Telescope (SOHO/EIT; \citealt{Delaboudiniere1995}), NSO/SOLIS \HeI{} observations, the NSO/KPVT \HeI{} archive \citep{Livingston1976}, and Yohkoh/Soft X-ray Telescope (SXT) observations \citep{Tsuneta1991}. SOHO is a project of international cooperation between ESA and NASA. SDO is a mission of NASA's Living With a Star Program. We also used the homogeneous EUV synoptic-map dataset developed by the University of Oulu, which provides Carrington-rotation synoptic maps reconstructed from SOHO/EIT and SDO/AIA full-disk observations \citep{Hamada2020}. The Oulu dataset is described as a homogeneous EUV synoptic-map dataset based on SOHO/EIT and SDO/AIA observations, covering SOHO/EIT maps from 1996--2018 and SDO/AIA maps from 2010--2018, and designed for long-term studies of coronal structures such as coronal holes \citep{Hamada2020}. We further acknowledge the use of the International Sunspot Number from SILSO, the World Data Center SILSO at the Royal Observatory of Belgium \citep{SILSO_Sunspot_Number}, and the F10.7~cm solar radio flux data provided by NOAA's Space Weather Prediction Center \citep{NOAA_SWPC_F107}. This work also made use of open-source scientific Python software, including SunPy \citep{SunPy2020}, NumPy \citep{Harris2020}, OpenCV \citep{Bradski2000}, Matplotlib \citep{Hunter2007}, Pillow, and PyTorch \citep{Paszke2019}.

\bibliography{aia_generation}
\bibliographystyle{aasjournalv7}
\end{document}